\documentclass[%
 aip,
 jmp,%
 amsmath,amssymb,
 reprint,%
]{revtex4-1}

\usepackage{graphicx}
\usepackage{dcolumn}
\usepackage{bm}

\begin{document}

\preprint{APS/123-QED}

\title{Flexible control of the Peierls transition in metallic C$_{60}$ polymers}
\author{Shota Ono}
\email{shota-o@eng.hokudai.ac.jp}
\author{Hiroyuki Shima}%
\affiliation{%
Division of Applied Physics, Faculty of Engineering,
Hokkaido University, Sapporo, Hokkaido 060-8628, Japan
}%

\date{\today}

\begin{abstract}
The metal-semiconductor transition of peanut-shaped fullerene (C$_{60}$) polymers is clarified by considering the electron-phonon coupling in the uneven structure of the polymers. We established a theory that accounts for the transition temperature $T_c$ reported in a recent experiment and also suggests that $T_c$ is considerably lowered by electron doping or prolonged irradiation during synthesis. The decrease in $T_c$ is an appealing phenomenon with regard to realizing high-conductivity C$_{60}$-based nanowires even at low temperatures.
\end{abstract}


\maketitle

With the developments in nanotechnology in the last decades, nano-carbon materials are regarded highly significant for potential electronic device applications. In particular, peanut-shaped fullerene (C$_{60}$) polymers are promising materials for application in nanometric conducting devices.\cite{onoe1,onoe2} C$_{60}$ polymers, produced through electron-beam irradiation (EBI) of pristine C$_{60}$ films, possess a quasi one-dimensional (1D) hollow tube geometry with a periodically modulated radius. It is a sort of $\pi$-electron conjugated systems, having metallic nature with the resistivity of 1-10 $\Omega$cm at room temperature.\cite{onoe1} A practical advantage of the synthesis is that it requires only EBI-induced polymerization of C$_{60}$ films deposited by dry (evaporation) or wet (spin coat) processes; hence, metallic C$_{60}$-based nanowires can be easily fabricated within a specified area on any substrate. This feature indicates the superiority of C$_{60}$ polymer applications to electronic devices through conventional integration techniques.

The uneven structure and reduced dimensionality of the C$_{60}$ polymers strongly affect their electronic and phononic properties at a low temperature. For instance, their peanut-like curved shape results in a significant increase in the Tomonaga-Luttinger exponent, which describes the collective excitations of electrons in quasi-1D systems.\cite{shima1} In the case of phononic excitations, infrared spectral measurements evidenced a rapid growth of specific eigenmode peaks with the increase in the EBI time;\cite{takashima} this peak growth is attributed to the anomaly in the phonon density of states peculiar to quasi-1D systems.\cite{onoe3}

In contrast to the well-understood effects mentioned above, the effect of electron-phonon (e-ph) couplings on the low-temperature physics of the C$_{60}$ polymers remains to be explored. An important problem to be considered is the possibility of a coupling-induced metal-semiconductor transition (i.e., the Peierls transition) in the system. Recently, through pump-probe spectroscopy, an energy-gap formation was found in the conduction band of C$_{60}$ polymers below 60 K; this energy-gap formation suggested the Peierls transition, driven by e-ph couplings.\cite{toda} However, theoretical understanding of the transition is still inadequate. From the perspective of device applications, it is fundamental to know why the conducting nature of the C$_{60}$ polymers vanishes across the transition temperature, $T^{\mathrm{exp}}_{c}\sim 60$ K, and how the e-ph coupling in the uneven structure contributes to the transition.

In this Letter, we provide a theoretical interpretation of the observed Peierls transition in the C$_{60}$ polymers by considering the structure-property relationship of the polymers. The theory leads us to an artificial manipulation of the transition temperature by alkali doping or a prolonged EBI procedure, whose realizations are directly linked to developing the potential of the C$_{60}$ polymers as conducting nanowires.

\begin{figure}[b]
\center
\includegraphics[scale=0.5,clip]{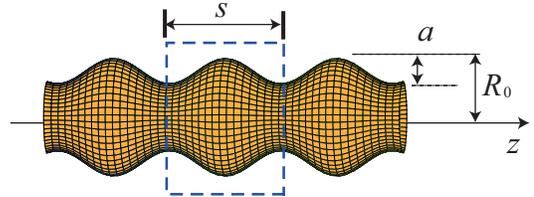}
\caption{\label{fig:tube}(Color online)
Continuum model of a 1D peanut-shaped C$_{60}$ polymer. The geometric values $R_0, a$, and $s$ are drawn. The square depicted by the dashed line (blue) indicates the unit cell.}
\end{figure}
For making an analytic argument, we map the C$_{60}$ polymer to a continuum thin hollow cylinder whose radius $R(z)$ is modulated along the axial $(z)$ direction as $R(z)=R_0 - a\sin^2(\pi z/s)$; we set $R_0=3.5$~\AA, $a=1.5$~\AA, and $s=7$~\AA \ (see Fig.~\ref{fig:tube}). The radius modulation yields two effective potential fields, each of which affects the motion of the electrons \cite{shima1,ono1} or phonons \cite{ono2} confined in the system. Due to the structural symmetry, a set of three quantum numbers $P=(k,\gamma,H)$ is required to specify the electronic states; here, $k$ is the electron's wavenumber, $\gamma$ indicates the angular momentum, and $H$ labels the electronic band.\cite{shima1} In the same manner, the phononic states are classified by a three-number set $J=(q,m,M)$, where $q$ and $m$ are the phonon wavenumber and angular momentum, respectively, and $M$ is the phononic band index.\cite{ono2}

\begin{figure}[tt]
\center
\includegraphics[scale=0.42,clip]{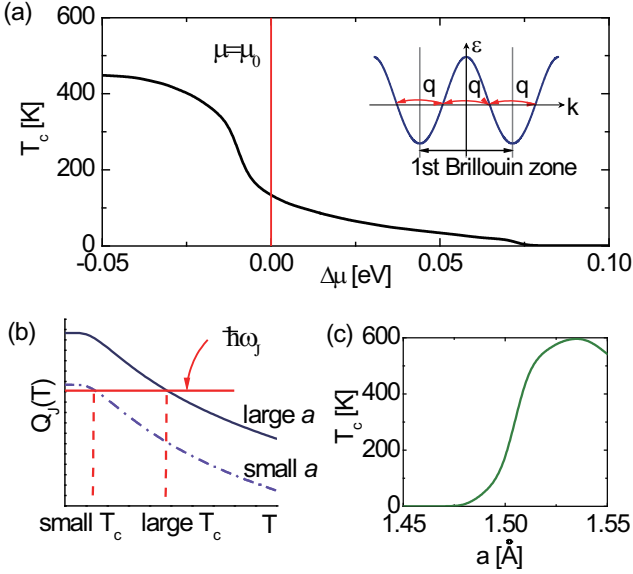}
\caption{\label{fig:temp_ef}(Color online)
(a) Transition temperature $T_{c}$ as a function of the Fermi level shift $\Delta \mu$ due to carrier doping. Without doping, we have $T_c\sim$130 K as indicated by a vertical line. Inset: Schematic of the commensurate effect. This schematic shows the electron band structure and relevant electron-phonon couplings via the phonon wavenumber $q$ that enhance $S_{P_i,P_f}$ [see Eq.~(\ref{eq:temperature2})] near the half-filling state. (b) Downward shift of the $Q_J$ curve with decreasing $a$. The intersection of the $Q_J$ curve and the horizontal line with $\hbar\omega_J$ gives the $T_c$. (c) Dependence of $T_c$ on the radius modulation amplitude $a$.}
\end{figure}
We define $T_{c}$ as the temperature at which a renormalized phonon frequency becomes zero. Let us suppose that the eigenfrequency $\omega_J$ of bare phonons are renormalized into $\omega$ by the e-ph interaction. Then, the phonon Green's function can be expressed as
\begin{equation}
 D(J,i\nu_l)=\frac{2\hbar\omega_{J}}
 {(i\nu_{l})^{2}-(\hbar \omega_{J})^2-2\hbar\omega_J\Pi(J,i\nu_l)},
 \label{eq:phonon_green}
\end{equation}
where $\Pi(J,i\nu_l)$ is the self-energy due to the interaction within the random phase approximation (RPA), and $\nu_l\equiv 2\pi l k_BT$ ($k_B$ being the Boltzmann constant) with integer $l$ is the Matsubara frequency for bosons.\cite{abri} By carrying out the analytic continuation $i\nu_l \rightarrow \hbar \omega +i0$, we obtain the retarded Green's function $D(J,\omega)$, whose pole gives the renormalized phonon frequency $\omega$:
\begin{equation}
 (\hbar \omega)^2-(\hbar \omega_{J})^2-2\hbar \omega_{J}
 \mathrm{Re} \left[\Pi (J,\hbar \omega +i0)\right]=0.
 \label{eq:phonon_renormalized}
\end{equation}
Taking the limit $\omega \rightarrow 0$ in Eq.~(\ref{eq:phonon_renormalized}), we obtain the following equation, which determines $T_c$:
\begin{eqnarray}
 \hbar \omega_{J} &=& Q_J(T_c) \equiv 4\sum_{P_i}\sum_{P_f} \vert g_{J}^{P_i,P_f} \vert^2 S_{P_i,P_f}(T_c), 
 \label{eq:temperature1}\\
S_{P_i,P_f} (T) &=& - \frac{f(\epsilon_{P_f}-\mu,T)-f(\epsilon_{P_i}-\mu,T)}
 {\epsilon_{P_f}-\epsilon_{P_i}}.
 \label{eq:temperature2}
\end{eqnarray}
Here, $g_{J}^{P_i,P_f} \propto \int \psi_{P_f}^{*} \psi_{P_i} \mathrm{div} \bm{u}_J dS$ is the coupling strength of two electrons, designated by $\psi_{P_i}$ and $\psi_{P_f}$, in the case where the coupling is mediated by a phonon whose displacement vector is given by $\bm{u}_J$; the integral $\int dS$ is carried out over the entire surface of the modulated cylinder depicted in Fig.~\ref{fig:tube}.\cite{ono2} In Eq.~(\ref{eq:temperature2}), $f$ is the Fermi distribution function, $\epsilon_{P_i}$ is the electron energy, and $\mu\equiv \mu_0 +\Delta \mu$ is the Fermi energy, with $\Delta \mu$ representing the deviation from the initial value $\mu_0$ by carrier doping. To solve Eq.~(\ref{eq:temperature1}) with respect to $T_c$, we define $\mu_0$ such that 60 electrons per a unit cell occupy single-particle states from the bottom of the electronic band and the conduction band width equals 1 eV from the {\it ab initio} calculations.\cite{wang}

\begin{figure}[ttt]
\center
\includegraphics[scale=0.4,clip]{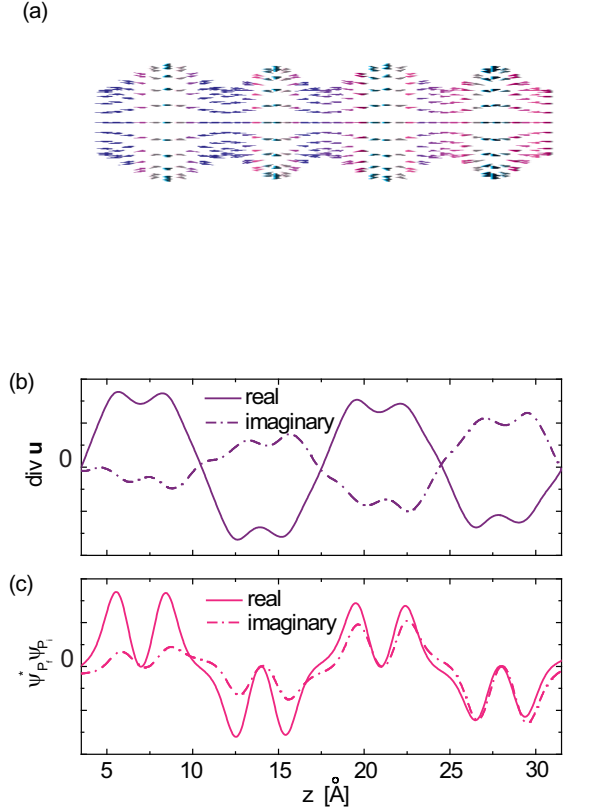}
\caption{\label{fig:distortion}(Color online)
(a) Displacement distribution of the phonon mode that mainly contributes to the Peierls transition under $\Delta \mu=0$ and $a=$1.5~\AA. (b) Profile of $\mathrm{div} \bm{u}_J$ of the phonon mode, shown in the plot of (a). (c) Profile of the wavefunction product, $\psi_{P_f}^{*} \psi_{P_i}$, of two electrons coupled via the phonon shown in (a).}
\end{figure}
Figure \ref{fig:temp_ef}(a) shows the $\mu$-dependence of $T_{c}$; see Fig.~\ref{fig:distortion}(a) for the relevant phonon mode at $\Delta \mu =0$.\cite{phonon_mode} We observe that $T_{c}$ decreases monotonically as $\mu$ increases; this decrease results from a reduction in the commensurate effect \cite{leung,gruner} that maximizes $T_c$ at the half-filling state [i.e., $\Delta \mu \sim -50$ meV in the present condition; see inset of Fig.~\ref{fig:temp_ef}(a)]. The most important finding from Fig.~\ref{fig:temp_ef}(a) is that we have $T_c=130$ K at $\mu=\mu_0$ (i.e., without doping). This result is seemingly at variance with the experimental data of $T_{c}^{\mathrm{exp}} \sim 60$ K; the deviation is possibly because actual C$_{60}$ polymers involve three-dimensional effects such as inter-chain electron hopping and Coulomb interaction between chains, which we have omitted in the RPA calculation. In fact, the relation $T_{c}^{\mathrm{exp}}=AT_{c}$ for $1/4 \lesssim A \lesssim 1/2$ has been commonly observed for many quasi-1D materials.\cite{toombs,gruner} Thus, we conclude that the energy-gap formation of the C$_{60}$ polymers below 60 K can be explained by the e-ph coupling theory, even in a quantitative manner, within the accuracy of the RPA. This is the first main result of the present work.

The rapid decrease in $T_c$ with the increase in $\mu$ is a significant phenomenon with regard to controlling the low-temperature electronic conductivity of the C$_{60}$ polymers. For example, we have found that an increment $\Delta \mu=$0.3 eV causes a drastic reduction in $T_c$ from 130 K to 1 K; this value of $\Delta \mu$ coincides with the upward shift in the Fermi level that is observed when one additional alkali atom per C$_{60}$ molecule is inserted into the hollow cavity of the C$_{60}$ polymer. As a result, the metallic nature of the C$_{60}$ polymers can survive even below 60 K, unlike the case of the pristine (undoped) system.\cite{experiment} This phenomenon provides a new avenue for artificially controlling the low-temperature conductivity of C$_{60}$ polymers, which is highly advantageous in actual device applications. This is the second result of this Letter.

Tuning the radial modulation amplitude (i.e., $a$) is another important technique for maintaining the metallic conductivity of the undoped C$_{60}$ polymers at temperatures below 60 K. It was experimentally found that $a$ decreases when the EBI in the synthesis is prolonged.\cite{onoe2} We suggest that such an EBI-induced reduction in $a$ can be utilized to reduce $T_c$, as conjectured from the $a$-dependence of the curve of $Q_J(T)$ appearing in Eq.~(\ref{eq:temperature1}). Figure~\ref{fig:temp_ef}(b) illustrates how a reduction in $a$ causes a downward shift in the curve of $Q_J(T)$ followed by a decrease in $T_c$; here, we reiterate that $T_c$ is defined as the temperature at which the curve of $Q_J(T)$ and the horizontal line of $\hbar\omega_J$ intersect. In fact, we have numerically confirmed from the plot of $T_c$. vs. $a$ in Fig.~\ref{fig:temp_ef}(c) that 2\% reduction in $a$ results in a drastic decrease in $T_c$, of more than two orders of magnitude. Therefore, well-controlled EBI time may serve as a tool for suppressing the Peierls transition, which endows the C$_{60}$ polymers with low resistivity at low temperatures. This is the third result of this study.

It follows from Eq.~(\ref{eq:temperature1}) that $T_c$ strongly depends on the e-ph coupling strength $g_{J}^{P_i,P_f}$ as well as $S_{P_i,P_f}$. For a better understanding of the e-ph coupling contribution to $T_c$, Fig.~\ref{fig:distortion}(a) illustrates the phonon mode that plays a dominant role in the occurrence of the Peierls transition at $\Delta \mu=0$ and $a=$1.5~\AA. Since the radius modulation yields an effective potential field with the same period of $s=$7~\AA, the resulting $\mathrm{div} \bm{u}_J$ becomes significant around the swelling region, as shown in Fig.~\ref{fig:distortion}(b). Similarly, the spatial profile of $\psi_{P_f}^{*}\psi_{P_i}$ at the Fermi level has a large amplitude in the vicinity of the swelling region [see Fig.~\ref{fig:distortion}(c)], owing to the curvature-induced potential effect.\cite{shima1} The synchronized oscillation of $\mathrm{div} \bm{u}_J$ and $\psi_{P_f}^{*}\psi_{P_i}$ results in an enhancement of $g_{J}^{P_i,P_f}$ (or equivalently, $Q_J$) in Eq.~(\ref{eq:temperature1}), which is sufficient to achieve the Peierls transition at an experimentally feasible temperature range. Artificial variances in $\Delta \mu$ and $a$ cause changes in the profiles of $\mathrm{div} \bm{u}_J$ and $\psi_{P_f}^{*}\psi_{P_i}$ of the relevant modes, thus yielding a change in the associated $T_c$, as demonstrated in the present work.

In conclusion, we have clarified the microscopic mechanism of the Peierls transition in C$_{60}$ polymers. The estimated transition temperature is quantitatively consistent with the experimental result.\cite{toda} Furthermore, we suggest that carrier doping and/or EBI time manipulation enable a significant reduction of the low-temperature resistivity of the C$_{60}$ polymers, which is an appealing feature with regard to C$_{60}$-based nanodevice fabrication.

We thank K. Yakubo, N. Nishiguchi, S. Mizuno, Y. Tanaka, Y. Toda, and H. Suzuura for valuable discussions. This study was supported by a Grant-in-Aid for Scientific Research from the MEXT, Japan. HS acknowledges the financial supports from the Inamori Foundation and the Sumitomo Foundation. Numerical simulations were carried out using the facilities of the Supercomputer Center, ISSP, University of Tokyo.

\nocite{*}



\begin{thebibliography}{99}
\bibitem{phonon_mode} In fact, only specific phonon mode with $J$ that gives the highest $T_c$ under a given $\Delta \mu$ should be relevant to the Peierls transition observed in practical measurements.
\bibitem{experiment} A similar reduction in the Peierls transition temperature by alkali doping has been reported in In atomic wires.\cite{hyungjoon}
\end{thebibliography}
\end{document}